# Promoting Astronomy Education: The Helix Nebula and Interdisciplinary Image Reading

**Fabiene B. Silva -** 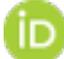 0009-0000-5509-7776

Physicist and Master's graduate student candidate for PCM (Science and Mathematics Education Graduate Program) at UEM (Maringá's State University)

**Vinicius Sanches -** 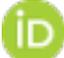 0009-0000-4996-2903

Biologist and Master's graduate student candidate for PCM (Science and Mathematics Education Graduate Program) at UEM (Maringá's State University)

## Abstract

The observation of space seems to have always caused wonder into people's collective consciousness, generating a series of historical myths. More recently specially with the development of better tools alongside the constant refinement of the scientific method Astronomy has consolidated into increasing field of Physics. Yet, representing such field in an accurate manner for beginner students poses a challenge. Appropriate images and descriptions should be chosen, which proves itself a large part of such challenge. Here we perform a technique named Interdisciplinary Image Reading aimed at trying to minimize the problem by improving and therefore promoting better Astronomy Education.

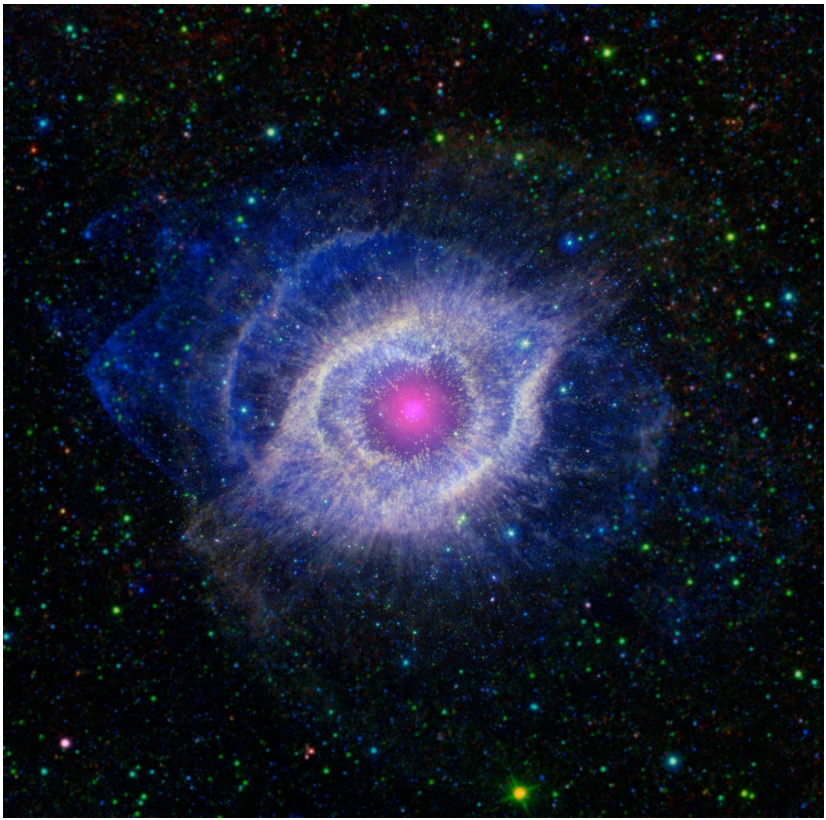

Fig. 1. Helix Nebula: Unraveling at the Seams (NASA & JPL-Caltech, 2012)

## 1. Introduction

Astronomy is a topic which is often featured in the media and seems to induce feelings of awe and curiosity in people. As of the nature of our increasing access to better graphics and better imaging techniques there exists a growing number of images, many of which are produced through post-production computational photography.

It would seem natural to assume a larger number of images would yield a better understanding of the field of Astronomy, but such assumptions are not always true. Although they should help learning, a careful selection of the images used have to be taken into consideration as they may also promote the development of alternative concepts which would not be desirable. In education the ideal role of images should be helping students form better mental representations and abstractions of the studied subject. To assure this goal and minimize undesirable alternatives concepts, the selection of appropriate images alongside detailed well written descriptions are crucial (Pena & Gil Quilez, 2001).

Here we perform a technique proposed by Silva and Neves (2021) named *Interdisciplinary Image Reading*, which divided in four steps aims to produce and predict objective and plausible subjective interpretations, or as called by the authors "readings", from images that we believe could be used to better select and describe the images used in Astronomy Education. To perform such technique the image labeled Figure 1, depicting the Helix Nebula, was selected and analyzed.

The image was chosen for two main reasons: The first one concerns the fact that it presents great potential within the Art and Science approach, enabling discussions to be held both around its scientific and artistic nature. The second reason relates to its aesthetics, considering the fact that it is an image that attracts attention, might awake different feelings from its observers and is also capable of providing a series of deeper questions and interpretations, going beyond a merely objective analysis.

## 2. Interdisciplinary Image Reading - The chosen image and the four steps of *Interdisciplinary Image Reading*

The Helix Nebula in a composite, computational photography image made by rendering data from the Spitzer Space Telescope and GALEX (Galaxy Evolution Explorer), when loaned to the California Institute of Technology in Pasadena, added to data captured from WISE (Wide-field Infrared Survey Explorer), all artificial satellites dedicated to space observation. The image was released on NASA's official website on October 4th, 2012.

### 2.1. Step 1: Shape Analysis

When observing the image, we can initially separate it into two focal points: the Helix Nebula in the center and the background environment made up of colored points. For a superficial look, this description is sufficient, but we only need to look at the image more closely and we will begin to notice a greater amount of details and peculiarities.

To describe this image, we will focus our attention first on the outside and move towards the center. The part closest to the edges of the image has several colored dots in mostly green, but also in other colors, such as blue dots, some orange, red and pink. Regardless of the color, some dots are larger than others, and regardless of the color or size, some shine more and others less. All are opaque.

Advancing our observation to the center of the image immediately diverts our attention to a different object. Breaking the pattern of colored dots of multiple sizes, colors and luminous intensity that continues to make up the entire background of the image, even up to its central region, the nebula stands out. A central point of intense brightness, but translucent, with pink color is purposely positioned in the center of the image. From this pink center we can observe countless very small dots, which resemble dust, arranged in a way that indicates movement, as if they had moved away from the center after some type of explosion, similar to memories of other explosions that we can observe with the naked eye. It's as if something exploded and is now rapidly expelling its parts into a space far from the center.

The nebula draws the observer's attention, not only because it is centrally positioned in the figure, but also because it is something very different from the pattern observed in its surroundings. Its shape resembles a human eye, its colors are brighter and its translucent pink center. We can deduce movement by observing something that resembles shiny dust appear to be moving away from this center and clumping together in layers of different densities, something we can associate with memories we have of explosions or impacts.

## 2.2. Step 2: Content Analysis

The image of the Helix Nebula that we are analyzing here can be framed within the broad theme of Astronomy and more specifically related to the area of space observation. A nebula or nebula consists of a large amount of dust and gases that act as "nurses" for new stars. They are formed by clouds of interstellar gases and dust or after a supernova - the explosion of a star (Space Center Houston, 2020).

Spatial images with greater detail and better resolution are captured in a more complex way than conventional photography. Satellites specialized in observing spectrums of light not visible to the eye collect this data, which is then compiled and rendered in colors that we can see.

To create this image, infrared data obtained by Spitzer from the center of the nebula were rendered in green and red. Data from WISE were used to assemble the outer area of the nebula and rendered in the same colors and ultraviolet data obtained from GALEX were rendered in blue (Dunbar, 2017).

Space observation is something that has probably always intrigued most human beings. The nature of the sky cyclically switching from a bright space during what we call "day" to a dark space containing different objects and phenomena that we call "night" is fundamentally peculiar compared to other changes we observe in everyday life. The advancement of Physics in conjunction with the advancement of technology constantly allows us to better understand the nature of reality and develop models that better explain its phenomena, something that is often linked to Astronomy in a two-way path - studying space advances Physics and Advances in Physics improve the study of space.

## 2.3 Step 3: Analysis of the Relationships Involving the Image 44(Author, Context and Reader)

The Helix Nebula, also known as NGC 7293, is a planetary-type nebula. It receives this classification due to its similarity to gas giant planets. Initially discovered in 1824 by Karl Ludwig Harding, a German astronomer, this nebula became one of the most famous, arousing the curiosity of amateur and professional photographers.

Although it was discovered in a context of astronomical studies, it is currently very popular in the non-scientific community as well, being relatively close to us, which allows it to be viewed using suitable technological devices.

The Helix Nebula is also constantly used as an example to explain the life and evolution of stars. Even today it is the target of important research in the astronomical world, which allows us to understand the intrinsic characteristics of stars, such as their spectral properties, their chemical composition and their structure.

The study of this nebula also allows us to understand what the evolution process of the Sun, the central star of our solar system, will be like. We know that, just like the star that resulted in the Helix Nebula, our Sun will "burn" all its fuel, gradually cooling until it becomes a white dwarf and, during this process, the outer gas layers will be expelled, forming a nebula .

Therefore, we observed that the image analyzed is worthy of the fascination and curiosity it arouses, bringing scientific knowledge to multiple contexts.

### 2.3. Step 4: Reader Interpretive Analysis

In general, it is possible to say that the image represents a beautiful work of cosmic art, because in addition to its significance for Astronomy and Astrophysics, it still arouses people's fascination due to its multicolored and "poetic" nature. After all, the story behind it encourages us to think about the infinity of our Universe, as well as its formation and evolution.

Its distinct appearance, which resembles an eye, earning it nicknames such as "Eye of God" and "Eye of Sauron", reminds us of the idea of a Universe that is observing us, generating a dichotomy between the human being observer and observed.

Its bright and intense colors are capable of sharpening the senses of those who observe them, resulting in feelings of euphoria and well-being, as well as relaxation and comfort. However, it is one of the most sought after and studied nebulae, both by astronomers and people curious about our Universe.

## 3. Final Considerations

In this paper we analyzed an image of the Helix Nebula published in 2012 by NASA and demonstrated the four steps of the technique of *Interdisciplinary Image Reading* developed by Silva and Neves (2021). To describe in words something extremely rich in detail, and to analyze the content presented on it in a technical way is often a challenged that such method tries to simplify. At the same time we presented some of the Nebula's history, as well as a deeper exploration of the importance of understanding the meaning and potential for its use in Astronomy education.

It is worth noting that the analysis presented here also takes into account our personal subjectivity and, as such, can change depending if someone else would be performing the same procedures. Even though there are criteria that must be followed while performing such *Interdisciplinary Image Reading*, it leaves room for different interpretations, meaning that the same image can be analyzed yield slightly different results.

## 4. Cited References